\documentclass[a4paper,fleqn,usenatbib]{mnras}

\usepackage{graphicx}	
\usepackage{amsmath}	
\usepackage{amssymb}	
\usepackage{lscape}
\usepackage{textcomp}
\usepackage{multirow}
\usepackage{times}

\title[]{Multiple stellar populations in the globular cluster M3 (NGC~5272): a Str\"{o}mgren perspective}

\author[D.Massari et al.]{
Davide Massari,$^{1,2}$\thanks{E-mail: davide.massari@oabo.inaf.it}
Emilio Lapenna,$^{3,1}$
Angela Bragaglia,$^{1}$
Emanuele Dalessandro,$^{3}$
\newauthor  Rodrigo Contreras Ramos,$^{4,5}$
P{\'{\i}}a Amigo$^{4,5}$
\\
$^{1}$INAF-Osservatorio Astronomico di Bologna, via Ranzani 1, I$-$40127, Bologna, Italy\\
$^{2}$University of Groningen, Kapteyn Astron Institute, NL-9747 AD Groningen, Netherlands\\
$^{3}$Dipartimento di Fisica e Astronomia, Universit\`a degli Studi di Bologna, v.le Berti Pichat 6/2, I$-$40127 Bologna, Italy\\
$^{4}$Millennium Institute of Astrophysics, Av. Vicuña Mackenna 4860, 782-0436 Macul, Santiago, Chile\\
$^{5}$Instituto de Astrofísica, Pontificia Universidad Católica de Chile, Av. Vicuña Mackenna 4860, 782-0436 Macul, Chile\\
}

\date{Accepted 2016 March 8. Received 2016 March 4; in original form 2016 February 10}

\pubyear{2015}

\begin{document}
\label{firstpage}
\pagerange{\pageref{firstpage}--\pageref{lastpage}}
\maketitle

\begin{abstract}
We present Str\"{o}mgren photometry of the Galactic Globular Cluster M3 to study its multiple generations phenomenon. 
The use of different colour-magnitude diagrams and especially of the notoriously efficient $c_{y}$ index allowed us to detect a 
double Red Giant Branch in the cluster CMD. 
After decontamination from fore- and background sources, the two sequences turned out to be equally populated. 
The two components also show a bimodal radial distribution well corresponding to
that predicted by numerical simulations for clusters living in an intermediate dynamical evolutive state
and with a population with modified chemical composition that was born more centrally concentrated than the primordial.
The analysis of high-resolution spectra quantitatively demonstrates that the two detected sequences correspond
to the first (Na-poor) generation and the second (Na-rich) generation, thus confirming the importance of synergy
between photometry and spectroscopy.
\end{abstract}

\begin{keywords}
Hertzsprung-Russell and colour-magnitude diagrams - stars: Population II  - stars: abundances – techniques: photometric –  globular clusters: general
\end{keywords}



\section{Introduction}

Galactic Globular Clusters (GCs) had been considered the best example 
of simple stellar population (e.g., \citealt{renzini86}). However, in the last decades strong
observational evidence has been gathered to prove that this is only a first approximation. In fact, GCs host multiple populations
of stars differing in terms of chemical composition (for a recent 
review see \citealt{gratton12}) which also reflects in spreads and splits of the evolutionary sequences 
in colour-magnitude diagrams (CMDs, see e.g., \citealt{monelli, piotto15}). 

The multiple population phenomenon is very complex and not totally understood yet (see \citealt{salaris14, bastian15a, renzini15}). 
The most common chemical signature of GCs is the star-to-star dispersion in light elements (C, N, O, Na, Mg, Al), showing anti-correlations and bi- or multi-modality 
(see \cite{gratton12}). 
All the massive Milky Way GCs studied so far,\footnote{The only exception to date is Ruprecht 106 (\citealt{villanova13}), which is massive and relatively young. 
Terzan 8, a GC belonging to the Sgr dSph, shows a hint of Na enhancement in a minority of stars \citep{carretta14}, 
while Terzan~7 and Pal~12, another two Sgr clusters, do not show anti-correlations, albeit on very small samples \citep{sbordone04,taut,cohen04}.} 
show the Na-O anti-correlation (see \citealt{carretta09a} for an extensive database, \citealt{bragaglia15} and references therein) and some of them also show a 
Mg-Al anti-correlation (e.g., \citealt{carretta09b,apogee}).
Up to now, a sub-sample of properly studied (see \citealt{bastian15} and references therein) show evidence of internal helium variation, 
with extreme cases being $\omega$-Centauri (\citealt{bedin04, piotto05, dupree}) and NGC~2808 (\citealt{dantona05, piotto07, dalex11, pasquini11, milone15a, massari16}).

Much less common is instead an internal dispersion in iron content, often combined with a dispersion in neutron-capture elements. 
Apart from the extreme cases of $\omega$~Cen \citep{norris95, lee99,jp10,marinoomega} and Terzan~5 
\citep{f09, origlia11, o13, massari14a}, with variation larger than 1 dex, several other clusters show variations of (some) 0.1 dex at most.
M54 \citep{carretta10}, M19 \citep{johnson15} NGC~5286 \citep{marino15} and M2 (\citealt{yong14, milone15a}, but see also \citealt{lardo16} who
found the cluster to have only a very small iron-enriched population) belong to this class of object.
For other clusters like M22 (\citealt{marino11, muccia15a}), NGC3201 (\citealt{muccia15b}) and NGC1851 (\citealt{carretta1851, yonggru}) 
the existence of an intrinsic iron spread is still a matter of debate.
Since a spread in the iron content might only be explained in terms of supernovae
ejecta self-enrichment, for all these clusters a peculiar origin is usually invoked.

Photometric studies showed that
the GC multiple population phenomenon (in the following, we will refer to the terms population
and generation without distinction, since it is beyond the aim of this paper to favour multiple
population scenarios which support multiple episode of star formation as in \citealt{decressin, dercole}, 
or which do not, see \citealt{bastian13}) can be addressed in detail
using a proper combination of filters (see e.g. \citealt{han,carretta11,milone12,monelli}).
In this sense, one of the best way to study the multiple populations of GCs comes
from medium-band Str\"{o}mgren $uvby$ photometry.
In fact the $u$ and $v$ filters of this set cover the spectral region including
the NH and CN molecular bands at  3400\AA~ and  4316\AA, respectively, and
they are therefore sensitive to variations in the C and N abundances.
The proof of the efficiency of Str\"{o}mgren photometry for this kind of study
comes from both observational and theoretical works such as \cite{yong08,carretta11, sbordone11,roh}, 
where the authors built suitable combination
of filters to enhance the separation among multiple RGB sequences.

In this paper, we focus our attention on the GC M3 (NGC~5272).
M3 is a northern, metal-poor GC, with a metallicity of about [Fe/H]$\simeq-1.5$ dex (\citealt{harris}, 
2010 edition), which shows evidence of variation in Na, O, Mg and Al (\citealt{sneden04, johnson05, cm05,
apogee}). Moreover, the analysis of its HB morphology performed in \cite{dalex13} by using HST photometry 
inferred an internal helium spread of $\Delta$Y$\simeq 0.02$. 
This result has been recently qualitatively confirmed by \cite{valcarce16} using Str\"{o}mgren photometry. 
Until now, no high-precision photometric studies have been dedicated
to the search and analysis of multiple populations in this cluster. Only a rough evidence for an anomalous 
spread of its RGB colour distribution was found by \cite{lardo11} using Sloan photometry, while 
\cite{piotto15} simply showed the existence of a double RGB in the cluster CMD coming from the HST UV Legacy Survey 
of Galactic GC, postponing the analysis to future papers. 
For these reasons we exploited deep images taken with Str\"{o}mgren $uvby$ filters
to provide a complete study of the multiple generations of stars in M3 and complemented it
with the analysis of high-resolution spectra.
Str\"{o}mgren photometry for M3 has been already used by \cite{gru99}.
However, this photometry is not publicly available and has furthermore been acquired on a small field of view ($3.8\times3.8$ arcmin$^2$, with the NOT). 
Our study covers a much larger field of view and we will make the catalogue available through the CDS\footnote{http://cdsweb.u-strasbg.fr}.

The paper is organized as follows. In Sect.\ref{data} we provide all the details on the photometric 
dataset and its reduction, while in Sect.\ref{cmds} we present the results coming from such an analysis.
In Sect.\ref{spec} the complementary spectroscopic analysis is described and the results
obtained from the comparison of the two investigations are shown in Sect.\ref{res}.
Finally, we draw our conclusions in Sect.\ref{concl}.

\section{Photometric data and analysis}\label{data}

\begin{figure}
	\includegraphics[width=\columnwidth]{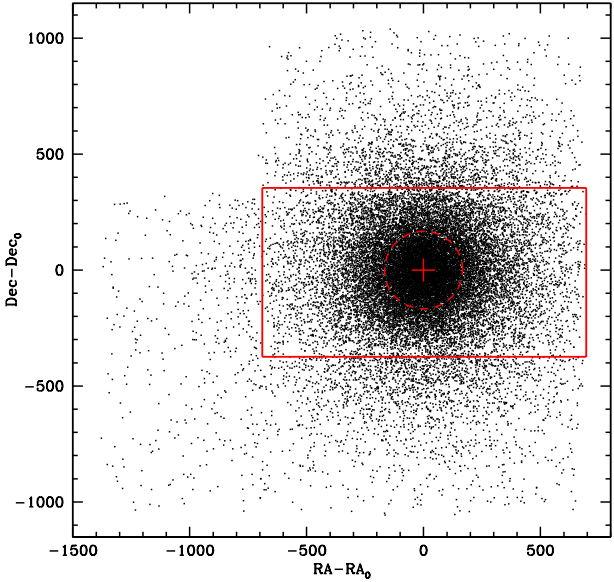}
        \caption{\small Map of the sources detected in our photometric catalogue. The centre of the cluster (\citealt{miocchi})
        is marked with a red cross. The size of chip 4, which is used throughout the analysis, is also delimited
        with a red box.}\label{map}
\end{figure}

The photometric dataset analysed in this work consists of a large sample of archival images
obtained with the Isaac Newton Telescope - Wide Field Camera (INT-WFC) in medium-band Str\"{o}mgren filters.
The INT-WFC detector consists of four chips, with a pixel scale of 0.33 \arcsec\,pixel$^{-1}$.
Each chip covers an area of $23.1\arcmin \times 12.1\arcmin$ and they describe a structured field of view 
where intra-chip gaps are about 1\arcmin~ wide, as shown in Fig.\ref{map}.

The retrieved archival observations belong to different datasets and cover several nights,
from the 22nd of March 2001 to the 9th of May 2006. During these nights
several dithered exposures of M3 have been taken in the $u,v,b,y$ Str\"{o}mgren filters.
Tab.\ref{tab:photo} summarises the entire dataset, listing the filters used, exposure times and the
total number of exposures analysed.

\begin{table}
	\centering
	\caption{\small Photometric dataset.}
	\label{tab:photo}
	\begin{tabular}{ccc} 
		\hline
		Filter & t$_{exp}$ [s] & n$_{exp}$\\
		\hline
		 \multirow{3}{*}{$u$} & 60 & 5 \\
                                      & 180 & 4\\
                                      & 500 & 1\\ 
		\hline     
		\multirow{3}{*}{$v$}  & 60 & 4 \\
                                      & 180 & 3\\
		\hline     
		\multirow{3}{*}{$b$}  & 10 & 8 \\
                                      & 15 & 1\\     
                                      & 20 & 1\\ 
                                      & 60 & 5\\ 
                                      & 180 & 2\\ 
                \hline     
		\multirow{3}{*}{$y$}  & 20 & 2\\ 
                                      & 60 & 5\\ 
                                      & 180 & 2\\    
                \hline     
	\end{tabular}
\end{table}

The pre-reduction of the raw images has been performed by means of standard procedures 
and the use of the IRAF\footnote{IRAF is 
distributed by the National Optical Astronomy Observatory, which
is operated by the Association of Universities for Research in Astronomy, Inc.,
under cooperative agreement with the National Science Foundation} package, treating each
chip of each exposure separately.
For each chip and filter, we retrieved from the Isaac Newton Group-ING archive
25 bias and 20 flat-field frames per filter, observed during the analysed observing runs.
We used these samples to compute $3$ sigma-clipped median bias and flat-field images by means of
the $zerocombine$ and $flatcombine$ IRAF tasks.
The final scientific images have been obtained using the $ccdproc$ IRAF task.

The photometric reduction has been performed by means of DAOPHOT and ALLSTAR (\citealt{stetson87}) 
and refined using ALLFRAME (\citealt{stetson94}). In
the latter step we used as input for ALLFRAME a catalogue of all the sources found in at least three $b$-band 
images, since they were those with the larger number of detections. After these steps, for each chip and filter 
we built a catalogue by requiring that each star fell in at least two of the reduced single exposures.
Since for each filter at least two short exposures exist, this choice allowed us
to always recover also the brightest stars that saturate in the long exposures.
We built a multi-wavelength catalogue for each chip by matching
the single-filter catalogues and including all the stars with at least two measured Str\"{o}mgren magnitudes.
Finally, for each single-chip catalogue we transformed stellar positions onto the 2MASS astrometric system (\citealt{2mass})
by using the CataPack software written by P. Montegriffo\footnote{{\tt www.bo.astro.it/\texttildelow paolo/Main/CataPack.html}}
and after calibrating their magnitudes we merged them together to build the final catalogue.
The calibration  onto the Str\"{o}mgren system has been achieved using 
ten standard stars observed during the night of the 22nd of March 2001.
Their calibrated magnitudes in the $uvby$ filters have been taken from the catalogue of
\cite{paunzen15}.

The spatial distribution of the sources that we recovered through our photometric
analysis is shown in Fig.\ref{map}, where all the stars are plotted as black dots. 
The centre of the cluster (RA$_{0}=13$:$42$:$11.38$, Dec$_{0}=+28$:$22$:$39.1$ from \citealt{miocchi}) 
is marked as reference with a red cross, while the red circle describes the cluster half-mass radius
(r$_{hm}$ $\sim167$\arcsec, see \citealt{miocchi}).

The aim of our work is to study the multiple stellar generations of M3. This requires high precision
in terms of photometric uncertainties. All the photometric standards have been observed only with the central 
chip 4 of the INT camera, and since only few stars are in common among the adjacent chips, we decided to focus 
the following analysis only on chip 4 itself. In this way, the uncertainties coming from the calibration procedure
among the four chips are avoided, and the covered field of view (see the red box in Fig.\ref{map}) 
is still sufficiently large to grant a good statistic in terms of detection. 
In fact, within the central chip fall $17155$ sources,
which correspond to the 79\% of the whole sample of sources detected considering the entire camera.
Moreover, the selected chip samples the central regions of M3 out to $\sim760$ \arcsec, that is more than
4.5 times the cluster r$_{hm}$.

\section{Str\"{o}mgren CMDs}\label{cmds}

\begin{figure}
	\includegraphics[width=\columnwidth]{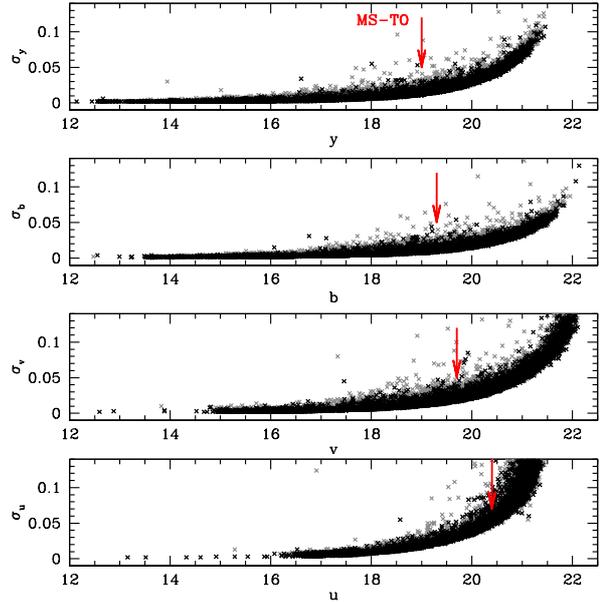}
        \caption{\small Photometric uncertainties for all the four Str\"{o}mgren filters used as a function of the 
        corresponding magnitude. Stars with $\mid sh\mid>0.2$ are plotted as grey dots. The location in magnitude 
        of the cluster Main-Sequence Turn-Off is marked with a red arrow.}\label{err}
\end{figure}

\begin{figure}
	\includegraphics[width=\columnwidth]{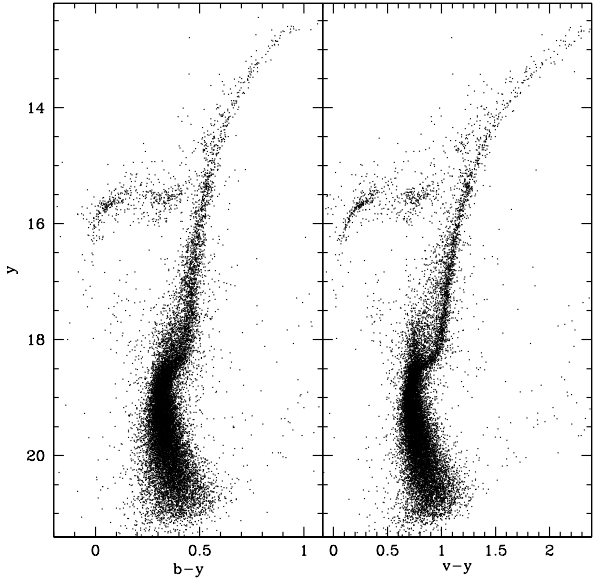}
        \caption{\small($y$, $b-y$) and ($y$, $v-y$) CMDs of M3.}\label{yby}
\end{figure}

The four panels of Fig.\ref{err} show the behaviour of the photometric errors $\sigma$, defined as the standard deviation of 
each star magnitude around the mean of the single-exposure measurements, as a function of the observed magnitudes. 
The scale of the plots has been kept fixed in order to show that in all filters, magnitudes brighter than the Main Sequence Turn Off (MS-TO,
marked in the Figure with red arrows) have internal uncertainties smaller than $0.03$ mag, the only exception being the u-band, 
where the median error at the MS-TO level is $0.05$ mag.
The ($y$, $b-y$) CMD of M3 is shown in the left panel of Fig.\ref{yby}. All the evolutionary sequences appear to be very well defined,
from a few magnitudes below the MS-TO to the RGB-tip, where stars saturate at $y<12.7$ mag. 
The same features are clearly visible in the ($y$, $v-y$) CMD (right panel of Fig.\ref{yby}), which covers a much wider colour
baseline.
A qualitative comparison with photometry presented in \cite{gru99} reveals no significant differences both in terms or the RGB 
extent and HB morphology.

\begin{figure}
	\includegraphics[width=\columnwidth]{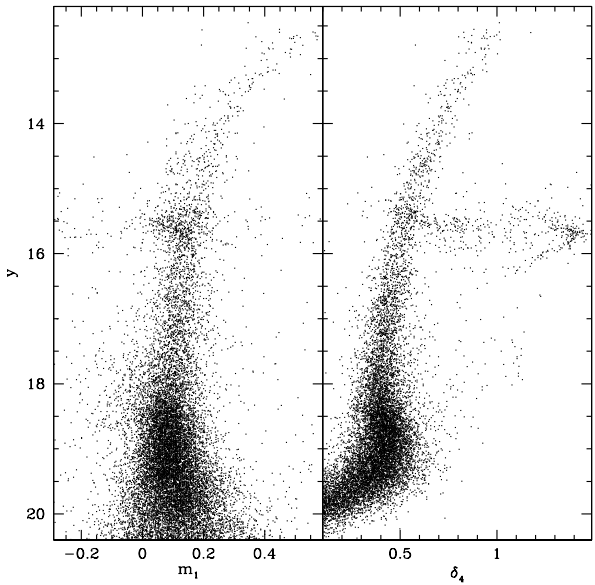}
        \caption{\small($y$, $m_1$) and ($y$, $\delta_4$) CMD of M3.}\label{m1d4}
\end{figure}

As shown in several previous works, a proper combination of Str\"{o}mgren filters can unveil the presence
of chemical anomalies in GC stellar populations. The index $m_1$ (defined as $m_1=$($v-b$)$-$($b-y$), \citealt{richter99}) is sensible
to the cluster metallicity. Left panel of Fig.\ref{m1d4} shows the ($y$, $m_1$) CMD of M3.
In this case, the spread of such index is $\sigma_{m1}=0.06$ mag at $16<y<16.5$, consistent with the photometric errors in this magnitude range and 
thus suggesting that M3 is homogeneous in terms of global metallicity.
On the other hand, the index $\delta_4=$($u-v$)$-$($b-y$) (\citealt{carretta11}) is able to reveal the presence of chemical
anomalies in the light element abundances of all the GCs studied so far, regardless of their global metal content. 
The right panel of Fig.\ref{m1d4} shows the cluster ($y$, $\delta_4$) CMD. The spread of $\delta_4$ at the same magnitude
interval as before is $\sigma_{d4}=0.08$ mag, larger than $\sigma_{m1}$.
Since such a value cannot be ascribed to larger photometric errors coming from the introduction of the $u$-band 
(that at the RGB level has uncertainties very similar to those measured for the other bands, see Fig.\ref{err}) in the index, 
we interpret it as suggestive of an intrinsic spread.

\cite{at95} used the index $c_1=$($u-v$)$-$($v-b$), which is sensible
to the strength of CN and CH bands. This colour index has been subsequently used for example by \cite{gru98} to study the
chemical anomalies in the stellar population of M13 or by \cite{yong08} for NGC~6752.
The last authors also defined another index called $c_{y}=$($c_1$)$-$($b-y$), which is still sensitive 
to N but insensitive to temperature, and which has been extensively used to distinguish between
first generation (FG) and second generation (SG) stars, being well correlated with Na abundances (see e.g. \citealt{carretta09a}).  
An extensive theoretical study on the influence of the typical chemical anomalies observed in GCs on the behaviour
of stellar sequences in Str\"{o}mgren CMDs has been presented in \cite{sbordone11}, while \cite{carretta11}
approached the problem from the observational side. According to both theoretical and observational
findings (see Fig.13 in \citealt{carretta11} to see the expected differences for stars N-rich and N-poor
in different evolutionary phases and of different metallicity), $c_{y}$ appears to be the most efficient index to study the multiple populations 
in M3 (which has a metallicity of about [Fe/H]$\sim-1.5$ dex).

\begin{figure}
	\includegraphics[width=\columnwidth]{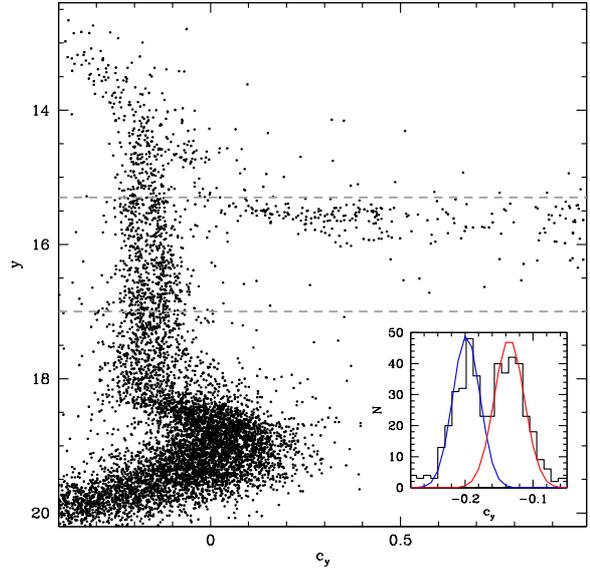}
        \caption{\small $c_{y},y$ CMD for stars with sharpness values $-0.2<sh<0.2$. The RGB clearly splits
        in two separated sequences. {\it Inset:} histogram of the colour distribution of stars in the RGB magnitude interval
        $15.3<$y$<17$.}\label{cyy}
\end{figure}

Fig.\ref{cyy} shows the ($y$,$c_{y}$) CMD, obtained after excluding all the sources with a sharpness parameter (defined by DAOPHOT 
as a shape-parameter whose value is zero for perfectly round and point-like sources as expected for stars) $\mid sh\mid>0.2$. They are 
also shown as grey dots in Fig.\ref{err}.

As immediately evident, in this combination of filters the RGB of M3 splits in two separated sequences.
The inset highlights the colour distribution of RGB stars in the magnitude range $15.3<y<17$, where the 
separation appears clearer. Such a distribution clearly displays two peaks, separated by about 0.05 mag.
According to \cite{carretta11}, the left sequence should be populated by N-poor, Na-poor FG stars,
while the right sequence by N-rich, Na-rich SG stars. 
Looking at their Figs. 1 and 2, we note that such a clear bimodal distribution seems rather unusual.
The only case resembling M3 is NGC~288, a slightly metal-richer and less massive GC. 
While the different level of N is demonstrated by photometry, we will come back on Na in Sections 
\ref{spec} and \ref{res}. 
To test the statistical significance of this bimodality, we performed a GMM test (\citealt{gmm}), which confirmed
that the unimodal hypothesis has to be rejected with a probability larger than 99.99\%. Such a result is confirmed 
independently also by a dip-test. The GMM test also gives as output the probability for each star to belong to 
one of the two components. 
We assigned each star to FG or SG (blue and red dots in Figure \ref{deco}, respectively) according to their membership probability.
In this way the right component accounts for 309 stars, 
that is the $52\pm3$\%, while the left component is made up of 283 stars,
corresponding to the $48\pm3$\%. Such an equal (within the uncertainty) distribution between FG and SG is quite rare among Galactic GCs 
and to date it has been observed only in NGC~6362 (see \citealt{dalex14})
and NGC~288 (\citealt{carretta11}) for which however there is some discrepancy between population     
ratios based on photometry and on spectroscopy (\citealt{carretta09a}).  Also NGC~2808 has an 
almost equal fraction of FG and SG stars, but this cluster presents more extreme behaviour and has 
recently been attributed no less than five separate populations (\citealt{milone15b, carretta15}).

\subsection{Radial distribution}

The radial distribution of multiple populations in GCs provides important information regarding their origin and formation (\citealt{vesperini13, dalex14, larsen15}).
According to the theoretical scenarios proposed so far, the second population(s) formed from the polluting material ejected by the 
primordial one should form in the central regions of a GC, thus appearing more centrally concentrated
(see e.g. \citealt{decressin, dercole, bastian15}). So far, most of the observations (e.g. \citealt{carretta09b, lardo11, beccari13})  
have confirmed this prediction. Few notable exceptions have been recently found, such as NGC6362 (\citealt{dalex14}) and NGC6121 (\citealt{nardiello15}), which do not show
significant radial differences at any distance from the cluster centres. This observational evidence can be interpreted as the effect of an efficient dynamical 
evolution possibly connected with a significant stellar mass loss (\citealt{vesperini13, dalex14, miholics15}).
Another unique case so far is represented by M15 
(\citealt{larsen15}), in which FG stars appear to be more centrally concentrated than SG ones. 
The authors suggest that this is due to peculiar initial conditions,
however alternative explanations have been recently proposed (\citealt{hb15}). 

\begin{figure}
	\includegraphics[width=\columnwidth]{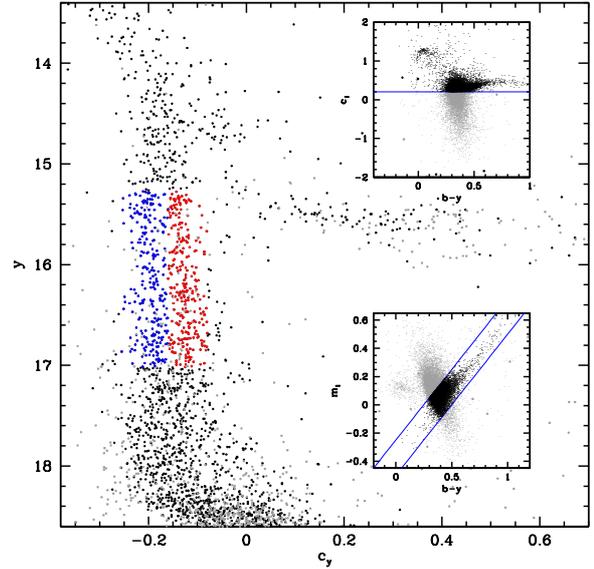}
        \caption{\small Decontaminated RGB of M3 (black dots) in the $c_{y}, y$ CMD. The selection criteria
        are shown in the two inset, where we followed the prescriptions in \citep{frank15} using the 
        $b-y, c_1$ and $b-y, m_1$ planes.}\label{deco}
\end{figure}

To study the radial distribution of multiple populations in M3, we focus on the RGB only. As a first step,
following the method described in \cite{frank15}, we decontaminated the samples of FG and SG RGB
stars described in Sect.\ref{cmds} using a proper selection in the $b-y, c_1$ and $b-y, m_1$ planes.
The insets of Fig.\ref{deco} show the criteria adopted for such a selection. In the upper inset,
we excluded in the $b-y, c_1$ all the sources with $c_1<0.2$ mag (see \citealt{arnadottir10}), while
in the lower inset we kept only stars running parallel to the iso-metallicity line of the
cluster RGB (see \citealt{calamida07}), within $\pm3\sigma$ from the best-fitting line.
After this procedure, 36 stars in the previously described magnitude interval ($15.3<$y$<17$)
were discarded, but the SG-to-FG ratio remained unchanged.

\begin{figure}
	\includegraphics[width=\columnwidth]{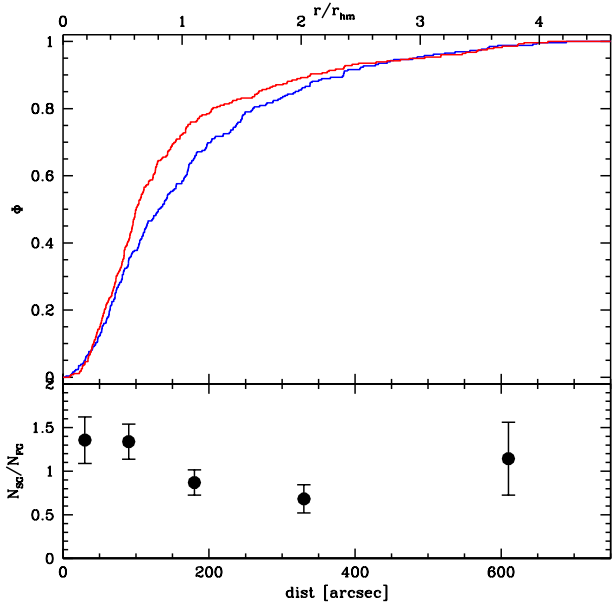}
        \caption{\small {\it Upper panel:} radial distributions for FG (blue) and SG (red lines) stars.
        {\ Lower panel:} SG-to-FG ratio as a function of the distance from the cluster centre. Radial bins
        are taken within 0.2, 0.6, 1, 2 and 3.5 r$_{hm}$.}\label{radial}
\end{figure}

At this point, in Fig.\ref{radial} we plot the radial distributions of the two decontaminated stellar generations. 
Overall, the trend observed for the red sequence is different from the other,
as demonstrated by a Kolmogorov-Smirnov (KS) test which gives a probability P$>99.8$\% that the two
distributions are not extracted from the same population. However, at a closer look this overall trend can be divided in
three separated intervals. 
Within the innermost 95\arcsec~ (i.e. $\sim0.6$ cluster r$_{hm}$ , see \citealt{miocchi}),
the two radial distributions appear very similar (the KS test gives a probability of being identical larger than 94\%)
and no gradient is found in the trend of SG-to-FG ratio against distance from the cluster centre (see
first two points at $0.2$ and $0.6$ r$_{hm}$ in the bottom panel of Fig.\ref{radial}).
On the other hand, outside $\sim0.6$ r$_{hm}$, the FG stars become progressively less concentrated but more numerous,
and the relative number ratio reaches a minimum at $\sim2$ r$_{hm}$. At that point, the trend inverts
and the SG-to-FG ratio reaches a value similar to that observed in the cluster central regions.
The final overall trend therefore describes a bimodal behaviour
and shows that the two populations are numerically similar only globally, but not locally.

This result demonstrates that it is crucial to cover the entire extent or a significant portion of a stellar cluster
to properly study the radial distribution of its multiple populations.

The SG-to-FG radial profile of M3 has been previously studied also by \cite{lardo11} based on Sloan photometry. 
Their sample did not cover the innermost regions of the cluster due to stellar crowding, and their analysis starts
at r$>100$\arcsec, where the two distributions appear to agree with our findings.
This peculiar bimodal profile resembles very well the prediction of \cite{vesperini13} for a cluster with an 
initially more concentrated SG and which underwent intermediate dynamical evolution (see their Figure 7). In this particular case,
the mixing of the two generations has started in the innermost regions, where the relaxation time is shorter due to the
larger number of interacting stars, but has not reached the entire cluster extent yet, as it happens in the most dynamically
evolved clusters (see \citealt{dalex14}). Such a dynamical state is also independently
confirmed by the radial distribution of Blue Straggler Stars, that in the dynamical clock framework described by \cite{ferraro12}
places this cluster among their Family-II sample (i.e. systems with intermediate dynamical age).

\section{Spectroscopic analysis}\label{spec}

By using Str\"{o}mgren photometry we have been able to separate two stellar populations
in the cluster M3. As already demonstrated in previous works (\citealt{carretta11, sbordone11}),
such a photometric feature indicates that the two populations are characterized by different chemical
abundances, in terms of light elements.
However, photometry can only give qualitative results in this respect. In order to 
quantitatively measure the difference in the light element abundances for the two populations
we need spectroscopy.
Several previous works in the literature studied the chemistry of M3 stars. Therefore
we looked for any spectroscopic target with previously measured abundances that has a counterpart 
in our photometric catalogue, and placed it on our CMD to test our findings.

We retrieved data published by \cite{sneden04} and \cite{cm05}, but
their targets are too bright and they either saturate or are non-linear in
our photometry.
From the list of \cite{johnson05}, we found 27 targets that in our photometry
are below the non-linearity limit (y$=14.1$ mag, anyway fainter than this value the sequences start to mix
together) and divided this sample into FG, Na-poor and SG, Na-rich stars depending 
on their Na abundance with respect to the median\footnote{This is not what 
\cite{carretta09a, carretta09b, carretta10} did. They based the separation between FG and SG on the minimum 
Na values in their samples of RGB stars, increased by 4 times the uncertainty on Na. However, 
the samples of stars we are using in the present paper are smaller and the adoption of a median 
value looks more robust.} [Na/Fe]$_{med}=-0.07$ dex.
\cite{johnson05} demonstrate that M3 shows not only
the common Na-O anti-correlation, but also that involving Mg and Al, which is not observed in all clusters
and tends to be more evident in metal-poor and/or massive GCs (e.g. \citealt{carretta09b, cordero15}).
For this reason, we exploited the recently published results of \cite{apogee},
where the authors measured [Al/Fe] abundances, and following the same selection criterium we divided their sample into
FG, Al-poor and SG, Al-rich stars, depending on each target Al abundance with respect 
to the median [Al/Fe]$_{med}=0.02$ dex.
The result of these selection is shown with triangles (Johnson's targets) and squares (M\'{e}sz\'{a}ros' targets) in Fig.\ref{targ}. 
Blue symbols represent FG targets, while red symbols show SG ones. RGB stars are marked with filled symbols,
while empty symbols highlight likely AGB and HB stars.
The spectroscopic separation nicely follows that achieved through photometry.
However, despite the good agreement, these targets do not sample the magnitude interval where the
photometric separation is more neat. Concerned by this fact, we looked for other targets that could fill
such a lack.

We used spectra acquired for a project on RR Lyrae stars in M3 with the multi-object spectrograph FLAMES (\citealt{pasquini02}) 
at the ESO-VLT (programme 093.D-0536, PI: Contreras Ramos), in a series of 7 exposures of $\sim$ 45 min each
with the HR12 setup (5821 - 6146\AA, R = 18700), which contains the Na D lines. Since our aim is to spectroscopically confirm whether the two RGB sequences we detected 
in the CMD of M3 correspond to FG and SG, respectively, we selected only the 17 spectra of the dataset belonging to RGB targets. 
Since these stars were observed as comparison for HB stars, their magnitude is fainter than for the 
literature samples and covers the region where the photometric separation is more evident.
The data reduction was performed using the dedicated ESO pipeline which includes bias
subtraction, flat fielding, wavelength calibration, and spectrum extraction. Individual stellar spectra have been cleaned 
from sky contribution by subtracting the corresponding median sky spectrum. Finally, multiple spectra of each target have been co-added, 
reaching a signal-to-noise (S/N) ratio per pixel at about 6000\AA~ of $\sim$ 50 in the faintest stars and
up to 80 in the brightest ones.

After the pre-reduction, we first measured the radial velocities (v$_{r}$) of each target to determine their membership to the cluster. 
Radial velocities have been obtained by using DAOSPEC (\citealt{daospec}) and by measuring the position of 34 lines distributed along the whole spectral range
covered by the HR12. The uncertainties have been computed as the dispersion of
the velocities measured from each line divided by the square root of the number of lines used, and they turned
out to be smaller than 0.4 km$\,$s$^{-1}$. The heliocentric corrections have been computed with the IRAF task RVCORRECT.
All the v$_{r}$ measurements and their uncertainties are summarised in Table \ref{tab:par}.
The mean velocity of the sample turned out to be v$_{r}=-148.8$ km$\,$s$^{-1}$  ($\sigma_{v}=3.4$ km$\,$s$^{-1}$), in very good agreement
with previous estimates (e.g. v$_{r}=-147.6$ km$\,$s$^{-1}$, \citealt{harris}).
Since all of the target velocities lie within $\pm3 \sigma_{v}$ from the mean velocity of the cluster (even if we adopt a more
robust estimate of $\sigma_{v}=5.6$ km$\,$s$^{-1}$, see \citealt{mc05}) we considered them all as cluster members.

Effective temperatures (T$_{eff}$) and gravities ($\log$ $g$) have been determined photometrically. Since not all of
our targets have a counterpart in the Str\"{o}mgren photometric catalogue (three are missing, see Tab.\ref{tab:par}), to
determine the atmospheric parameters we projected the location of each target in the optical (V, V-I) CMD onto a reference isochrone. 
This CMD is built from two separated catalogues. One comes from the publicly available HST photometry of \citealt{sarajedini} 
and samples the innermost regions of the cluster, while the other comes from archival MegaCam observations and samples the outer regions. 
We selected as isochrone an $\alpha$-enhanced model with age of 12 Gyr, Z$=0.001$ (corresponding to [Fe/H]$\simeq-1.6$ dex)
and standard helium composition retrieved from the BaSTI archive (\citealt{basti06}) and we placed it onto the CMD assuming
distance modulus and reddening by \cite{ferraro99}. 
The errors associated to these parameters have been computed as described in \cite{massari14b}, by taking into account
the uncertainties on the projection arising from distance modulus ($\sigma_{DM}=0.1$ mag), reddening ($\sigma_{E(B-V)}=0.05$ mag)
and photometry ($\sigma_{V}=\sigma_{I}=0.02$ mag). They turned out to be of the order of $\sim70$ K for T$_{eff}$ and 0.1 dex for $\log$ $g$.
Unfortunately, for four targets of our sample we were not able to determine a photometric counterpart, and they have been therefore excluded 
from the analysis.

The atmospheric parameters were derived by using photometric informations
because the limited wavelength range covered by the HR12 does not allow us to use a sufficiently
large number of FeI and FeII lines in order to use the fully spectroscopic approach based on the excitation/ionization balance.
We used the code GALA\footnote{http://www.cosmic-lab.eu/gala/gala.php} (\citealt{gala}) to determine the microturbulent velocity (v$_{turb}$) by requiring
that no trend exists between iron abundances and line strengths of the lines used,
while the photometric temperatures and gravities have been kept fixed during the analysis.
For the microturbulent velocity we obtained an average uncertainty of about 0.2 km$\,$s$^{-1}$ .
The atmospheric parameters of the analysed targets are listed in Table \ref{tab:par}.

\subsection{Elemental abundances}

The abundances of iron have been computed by using GALA. 
The EW and the error of each line were obtained using DAOSPEC, iteratively launched by means of the 
4DAO\footnote{http://www.cosmic-lab.eu/4dao/4dao.php} code (\citealt{4dao}).
The lines considered in the analysis have been selected from suitable synthetic spectra at the FLAMES resolution and computed with
the SYNTHE package (\citealt{sbordone04}) by using the average parameters estimated from photometry and the metallicity
derived by \cite{carretta09a}.
The model atmospheres have been computed with the ATLAS9\footnote{http://wwwuser.oats.inaf.it/castelli/sources/atlas9codes.html} code (\citealt{atlas}).
We adopted the atomic and molecular data from the last release of the
Kurucz/Castelli compilation\footnote{http://wwwuser.oats.inaf.it/castelli/linelists.html} and selected
only the lines predicted to be unblended.
The final iron abundances of the targets have been computed as the average of each
single line measurement, by rejecting all the lines with an EW uncertainty larger than 20\%.
The errors have been computed by dividing the line-to-line dispersion by the square root of the number of lines used.
The average iron abundance for the whole sample of 17 giants is [Fe/H]$=-1.40$ dex ($\sigma=0.09$ dex).
Such a value is in good agreement with previous estimates like \citealt{carretta09a} or
\citealt{cavallo00}, which quote [Fe/H]$\simeq-1.5$ dex, and in excellent agreement with \cite{apogee},
who found [Fe/H]$=-1.40$ dex ($\sigma=0.08$ dex).

Abundances of NaI have been derived
with the spectral synthesis technique performed by using the package SALVADOR (Mucciarelli et al., in prep)
from the lines at 5886 and 5892$\rm\mathring{A}$.
The corresponding internal uncertainties have been computed by means of Monte Carlo simulations. 
In particular, we added Poissonian noise to the best-fit synthetic spectrum to reproduce the observed SNR and then 
we repeated the analysis.  After 1000 Monte Carlo realizations, the dispersion of the measurements
around the mean value has been adopted as the internal abundance uncertainty.
In order to take into account Non Local Thermodynamical Equilibrium (NLTE)
effects the abundances derived from the NaI lines have been corrected according to \cite{lind11b}.
The adopted solar reference value have been taken from \citet{grevesse98}.
The final values together with the related uncertainties are listed in Table \ref{tab:par}.

\section{Results}\label{res}

Fig.\ref{na_dist} shows the numerical distributions of [Na/Fe] with (upper panel) and without (lower panel)
correction for NLTE effects.
Their dispersions are $\sigma_{[Na/Fe]}^{LTE}=0.19$ dex and $\sigma_{[Na/Fe]}^{NLTE}=0.24$ dex, respectively. These values
agree well with the finding by \cite{johnson05}, who found $\sigma_{[Na/Fe]}^{LTE}=0.25$ dex and $\sigma_{[Na/Fe]}^{NLTE}=0.26$ dex
for a larger sample of 77 stars.

\begin{figure}
    \includegraphics[width=\columnwidth]{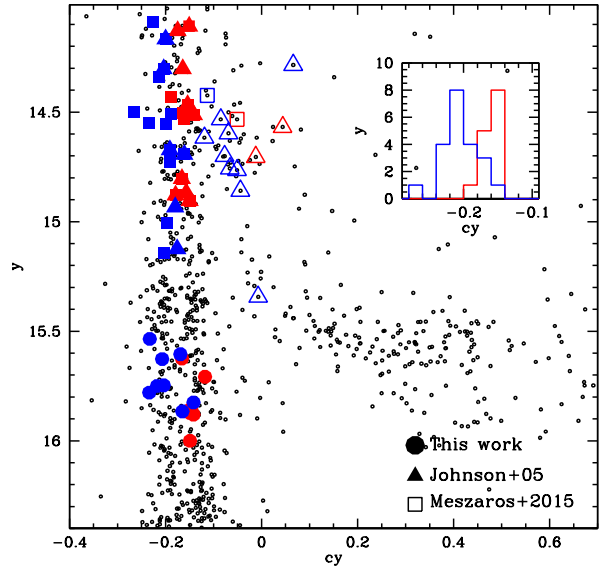}
        \caption{\small ($y$,$c_{y}$) CMD of M3 zoomed in the RGB region. Targets analysed in this work are
        highlighted as filled circles (blue ones corresponding to Na-poor, FG stars, red ones to Na-rich, SG), while
        targets taken from \citep{johnson15} are marked as filled triangles and those
        from \citep{apogee} as filled squares (with the same colour code). Empty symbols mark non-RGB stars.
        The inset shows the colour distribution of FG and SG RGB targets, revealing the same bimodality observed photometrically.}\label{targ}
\end{figure}

The two distributions do not show a clear-cut bimodality, but only a hint of a separation
in two groups. For these reason, 
we decided to assume the median values of the two distributions to split
the sample in Na-poor (FG) and Na-rich (SG) stars, as done for the stars retrieved from
the sample of \cite{apogee}. Median values are shown in Fig.\ref{na_dist} with
vertical red dashed lines.
We cross-correlated the spectroscopic targets with our photometric catalogue,
finding 14 matches (targets IDs, coordinates and magnitudes are listed in Tab.\ref{tab:par}, where ID$_{spec}$
corresponds to that used in the spectroscopic archive and ID$_{phot}$ to that used in our photometric catalogue.). 
Their location in the Str\"{o}mgren ($y,c_{y}$) CMD is shown
with filled blue (FG) and red (SG) circles in Fig.\ref{targ}. 

\begin{figure}
    \includegraphics[width=\columnwidth]{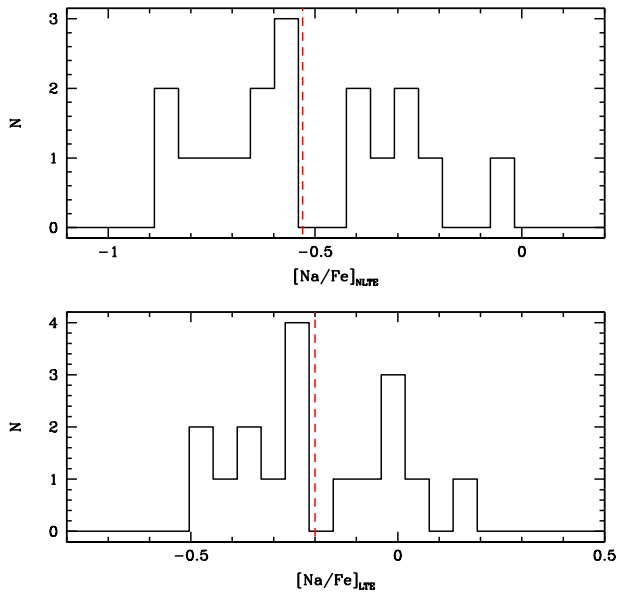}
        \caption{\small Histograms of the [Na/Fe] distributions of our analysed targets before ({\it lower panel}) and 
        after ({\it upper panel}) NLTE corrections. The median values of the two distributions are also
        marked with the two vertical red dashed lines.}\label{na_dist}
\end{figure}

The position of the targets follows quite well the bimodality of the RGB detected photometrically,
with all the Na-rich stars located on the right RGB, and the majority of Na-poor stars
distributed along the left RGB, with only few outliers.

In order to highlight the nice separation of all the spectroscopic FG and SG targets,
we show their colour distribution in the inset of the Figure. Two main
peaks appear to be very well defined, and this appearance is confirmed
by both the GMM test and the dip-test, which confirm the bimodality of the distribution
with a statistical significance larger than 99.9\%.
A further confirmation of the good correspondence between our photometric and
spectroscopic results comes from the number-ratio of the two populations.
In fact, from our spectroscopic sample we found that FG and SG
are equal in number within the Poissonian uncertainty, with the FG accounting 
for the $56\pm8$\% against the $44\pm11$\% of SG stars. This result is in good 
agreement with our photometric estimate ($48\pm3$\% of FG and $52\pm3$\% of SG 
members, see Sect.\ref{cmds}).
It is also worth noticing that among the 12 AGB targets found in the literature samples, 
the number ratio of the two population is significantly different from that measured in our sample
along the RGB. 
In fact, nine stars appear to belong to the FG and the remaining three to the SG, this leading
to a 75\%-25\% ratio. Such a finding is interesting because it represents
one of the few measurement in the disputed topic concerning the behaviour of GC multiple
generations along the AGB. In fact, recently a few metal-poor GCs were found to show no SG-AGB
stars at all. This is the case for NGC~6752 (\citealt{campbell13}) and for M62 (\citealt{lapenna15}),
though in the latter case the lack of SG-AGB stars is possibly due to the small sample of studied
stars. However, in several other cases (\citealt{johnson15b, garciah15}) SG-AGB stars have indeed been detected. 
M3 belongs to the second of these groups, since \cite{garciah15} already found three SG-AGB stars
using APOGEE spectra. In this work we confirm this finding and we add other three of such rare targets.

\section{Summary and Conclusions}\label{concl}

In this work, we exploited a ground based photometric dataset in Str\"{o}mgren
$uvby$ filters to study the multiple stellar generations of the GC M3.
The analysis of the CMD built with the $c_{y}$ index has shown the presence of
a double RGB, equally populated in its two components.
The radial distributions of the two populations turned out to be quite peculiar,
with red and blue sequence being completely mixed up to about the half-mass radius,
then changing the observed trend with the red component staying more centrally concentrated
out to about $\sim3.5$ r$_{hm}$, when the two mix again.
Such a trend is predicted to be observed in clusters which experienced moderate
dynamical evolution, and it confirms previous dynamical age estimate available
in literature.

The spectroscopic analysis of a sample of 17 giants observed with FLAMES, and the
comparison with previous abundance measurements available in literature, demonstrated
that the photometric detection of the two RGBs flags the presence of chemically distinct populations
with the FG corresponding to the left sequence and the SG to the 
right, as observed in all the other metal-poor and metal-intermediate clusters
studied so far with the Str\"{o}mgren $c_{y}$ index.
We also found three new SG-AGB stars, thus confirming that the lack of such stars observed in
other metal-poor GCs is not an ubiquitous feature.

Our work therefore demonstrates yet again how efficient Str\"{o}mgren photometry is to
study the phenomenon of multiple stellar generations in GCs over the large field of view
and the large number of samples stars permitted by ground-based photometric observations
and how important is to couple spectroscopic and photometric information to characterise
globular cluster populations.

\section*{Acknowledgements}

We thank the referee Judith Cohen for her report and suggestions which helped us to improve the presentation of our results.
This paper makes use of data obtained from the Isaac Newton Group Archive which is maintained as 
part of the CASU Astronomical Data Centre at the Institute of Astronomy, Cambridge.
This paper also uses observations made with ESO Telescopes at the La Silla Paranal Observatory 
under programme ID 093.D-0536.
DM has been supported by the FIRB 2013 (MIUR grant RBFR13J716).
EL and ED have been supported by COSMIC-LAB (web site: http://www.cosmic-lab.eu), which is funded by 
the European Research Council (under contract ERC-2010-AdG-267675).
AB acknowledges partial support from PRIN-MIUR 2010-2011  “The Chemical and Dynamical Evolution of 
the Milky Way and Local Group Galaxies” (PI F. Matteucci). 
PA acknowledges Proyecto Fondecyt Regular 1150345.
This research made use of the SIMBAD database, operated at CDS, Strasbourg, France and 
of NASA’s Astrophysical Data System.

\begin{landscape}
\begin{table}
	\centering
	\caption{\small Photometric and spectroscopic quantities of the analysed targets.}
	\label{tab:par}
	\begin{tabular}{rrccccccccccccc} 
		\hline
		ID$_{spec}$ & ID$_{phot}$ & R.A. & Dec. & u & v & b & y & v$_{r}$ & T$_{eff}$ & $\log$ $g$ & v$_{turb}$ & [Fe/H] & [Na/H]$_{LTE}$ & [Na/H]$_{NLTE}$\\
		\hline
		  &  & (J2000) & (J2000) & mag & mag & mag & mag & km$\,$s$^{-1}$ & K & dex & km$\,$s$^{-1}$ & dex & dex & dex \\
		\hline
		$100579$ & 1986 & 205.6562914 & 28.4329251 & 16.592 & 15.077 & 13.950 & 13.602 & $-151.9\pm0.3$ & $5068$ & $2.5$ & $1.85$ & $-1.42\pm0.09$ & $-1.42\pm0.08$ & $-1.72\pm0.17$  \\
		$103981$ & 6897 & 205.6844581 & 28.3859233 & 16.378 & 14.828 & 13.644 & 13.265 & $-148.7\pm0.3$ & $5019$ & $2.4$ & $2.05$ & $-1.32\pm0.10$ & $-1.58\pm0.03$ & $-1.91\pm0.10$  \\
		$ 18711$ & 1185 & 205.3888267 & 28.4484213 & 16.527 & 15.017 & 13.861 & 13.555 & $-149.1\pm0.3$ & $5060$ & $2.5$ & $2.05$ & $-1.43\pm0.13$ & $-1.66\pm0.02$ & $-2.01\pm0.07$  \\
		$ 25241$ &13795 & 205.4358151 & 28.3377528 & 16.607 & 15.074 & 13.945 & 13.584 & $-147.0\pm0.2$ & $5060$ & $2.5$ & $1.60$ & $-1.29\pm0.11$ & $-1.41\pm0.06$ & $-1.69\pm0.15$  \\
		$ 27510$ & 4189 & 205.4475563 & 28.4100074 & 16.565 & 15.068 & 13.875 & 13.596 & $-151.9\pm0.4$ & $5068$ & $2.5$ & $1.90$ & $-1.27\pm0.07$ & $-1.68\pm0.09$ & $-2.04\pm0.18$  \\
		$ 42796$ & 7302 & 205.5012666 & 28.3856410 & 16.462 & 14.985 & 13.796 & 13.480 & $-152.1\pm0.3$ & $5043$ & $2.4$ & $2.30$ & $-1.37\pm0.10$ & $-1.70\pm0.04$ & $-2.08\pm0.12$  \\
		$ 48341$ & 3543 & 205.5150179 & 28.4149743 & 16.432 & 14.993 & 13.812 & 13.510 & $-146.4\pm0.3$ & $5042$ & $2.4$ & $1.65$ & $-1.40\pm0.07$ & $-1.65\pm0.04$ & $-2.01\pm0.12$  \\
		$ 53478$ & 2477 & 205.5264945 & 28.4280449 & 16.378 & 14.846 & 13.650 & 13.335 & $-147.9\pm0.4$ & $5012$ & $2.4$ & $1.90$ & $-1.44\pm0.12$ & $-1.72\pm0.04$ & $-2.08\pm0.13$  \\
		$ 53570$ &14330 & 205.5267239 & 28.3307889 & 16.382 & 14.856 & 13.707 & 13.354 & $-148.0\pm0.2$ & $5011$ & $2.3$ & $1.85$ & $-1.41\pm0.09$ & $-1.50\pm0.05$ & $-1.82\pm0.14$  \\
		$ 67305$ &12304 & 205.5563429 & 28.3497774 & 16.486 & 14.926 & 13.783 & 13.438 & $-148.2\pm0.3$ & $5020$ & $2.4$ & $2.00$ & $-1.43\pm0.12$ & $-1.40\pm0.04$ & $-1.68\pm0.13$  \\
		$ 78766$ & 2615 & 205.5813052 & 28.4254582 & 16.663 & 15.176 & 14.040 & 13.730 & $-148.0\pm0.3$ & $5081$ & $2.5$ & $1.85$ & $-1.42\pm0.08$ & $-1.41\pm0.08$ & $-1.70\pm0.17$  \\
		$ 95001$ &  229 & 205.6299793 & 28.4645877 & 16.311 & 14.901 & 13.815 & 13.477 & $-148.1\pm0.4$ & $5043$ & $2.4$ & $1.95$ & $-1.70\pm0.11$ & $-2.08\pm0.16$ & $-2.50\pm0.19$  \\
		$ 96900$ & 8553 & 205.6379019 & 28.3754247 & 16.578 & 15.070 & 13.949 & 13.610 & $-140.3\pm0.3$ & $5057$ & $2.5$ & $2.05$ & $-1.42\pm0.10$ & $-1.25\pm0.04$ & $-1.49\pm0.11$  \\
		$ 97013$ &15559 & 205.6383057 & 28.3134164 & 16.328 & 14.851 & 13.716 & 13.357 & $-150.5\pm0.2$ & $5013$ & $2.4$ & $1.60$ & $-1.35\pm0.12$ & $-1.81\pm0.02$ & $-2.20\pm0.06$  \\
		$104765$ & - & - & - & - & - & - & - & $-157.5\pm0.5$ & $5023$ & $2.4$ & $1.90$ & $-1.41\pm0.10$ & $-1.87\pm0.08$ & $-2.28\pm0.15$  \\		
		$ 28686$ & - & - & - & - & - & - & - & $-148.5\pm0.3$ & $5048$ & $2.4$ & $1.95$ & $-1.45\pm0.09$ & $-1.46\pm0.03$ & $-1.77\pm0.08$  \\		
		$ 69664$ & - & - & - & - & - & - & - & $-149.0\pm0.2$ & $5073$ & $2.5$ & $1.50$ & $-1.33\pm0.07$ & $-1.58\pm0.05$ & $-1.92\pm0.10$  \\		
		\hline            
	\end{tabular}
\end{table}
\end{landscape}












\bsp	
\label{lastpage}
\end{document}